\documentclass[fleqn,twoside]{article}
\usepackage{espcrc2}
\usepackage{epsfig}

\usepackage{cite,amssymb,float}
\usepackage[usenames,dvips]{color}
\usepackage{amsmath, mathbbol}
\usepackage{graphicx}
\usepackage[figuresright]{rotating}

\title{Phenomenology of LFV at low-energies and at the LHC: strategies to 
probe the SUSY seesaw}

\author{Ana M. Teixeira\address{Laboratoire de Physique Corpusculaire, 
CNRS/IN2P3 -- UMR 6533,\\
Campus des C\'ezeaux, 24 Av. des Landais, F-63177 Aubi\`ere Cedex, 
France}
\thanks{Contribution to the Proceedings of 
``TAU 2010 - The 11th International Workshop on Tau Lepton Physics'', 
Manchester, UK, 13 - 17 September 2010.\newline
PCCF RI 1006}, 
Asmaa Abada\address{Laboratoire de Physique Th\'eorique, CNRS -- 
UMR 8627,\\ Universit\'e de Paris-Sud 11, F-91405 Orsay Cedex, France}, 
Ant\'onio J. R. Figueiredo\address[IST]{Centro de 
F\'{\i}sica Te\'orica de Part\'{\i}culas, 
Instituto Superior T\'ecnico, \\ Av. Rovisco Pais 1, 
1049-001 Lisboa, Portugal} and
Jorge C. Rom\~ao\addressmark[IST]}

\begin{document}

\begin{abstract}

We study the impact of a type-I SUSY seesaw concerning lepton flavour
violation (LFV) at low-energies and at the LHC. At the LHC, 
$ \chi_2^0\to \tilde \ell \,\ell \to  \ell \,\ell\,\chi_1^0$ decays,
in combination with other observables, render feasible
the reconstruction of the masses of the intermediate sleptons, and
hence the study of $\ell_i - \ell_j$ mass differences. If
interpreted as being due to the violation of lepton flavour,
high-energy observables, such as large slepton mass splittings and
flavour violating neutralino and slepton decays, 
are expected to be accompanied 
by low-energy manifestations of LFV such as radiative and three-body
lepton decays. We discuss how to devise strategies based in 
the interplay of slepton mass splittings as might be
observed at the LHC and low-energy LFV observables to derive
important information on the underlying mechanism of
LFV. {This contribution
summarises part of the work of Ref.~\cite{Abada:2010kj}}.  

\vspace{1pc}
\end{abstract}

\maketitle

\section{MOTIVATION}
Extending the Standard Model to accommodate $\nu$ data
naturally opens the possibility of many other
new phenomena. Embedding a type-I seesaw 
in supersymmetric (SUSY) models provides a unique framework where many
theoretical and experimentally shortcomings of the Standard Model can
be successfully addressed.
Within a type-I SUSY seesaw, even if the soft supersymmetry breaking
terms are flavour universal at some high energy 
unification scale, flavour violation appears at low-energies 
due to the renormalisation group (RG) evolution of the SUSY 
soft-breaking parameters, driven by the potentially large and
necessarily non-diagonal neutrino Yukawa
couplings~\cite{Borzumati:1986qx}.  
Low-energy manifestations of LFV 
in the framework of the SUSY seesaw include sizable branching ratios
(BR) for radiative decays as $l_i \to l_j \gamma$, three-body decays, $l_i \to 3
l_j$  and  sizable $\mu-e$ conversion rates (CR) 
in heavy nuclei (for a review, see 
Ref.~\cite{Raidal:2008jk} and references therein). 

At the LHC, processes such as slepton mediated neutralino decays 
$\chi_2^0 \to \ell_i^\pm\, \ell_j^\mp\, \chi_1^0$ offer a golden
laboratory to study lepton flavour violation at higher energies. 
Three possible signals of LFV can be observed at the LHC: (i)
flavoured slepton mass splittings, provided that one can effectively reconstruct
slepton masses via observables such as the kinematic end-point of the 
invariant mass distribution of the leptons coming from the above
mentioned cascade decay; (ii) multiple edges in di-lepton invariant mass
distributions $\chi_2^0 \to \chi_1^0\, \ell_i^{\pm}\, \ell_i^{\mp}$, 
arising from the
exchange of a different flavour slepton $\tilde l_j$ 
(in addition to the left- and right-handed sleptons, $\tilde
l_{L,R}^i$); (iii) sizable widths for LFV decay processes like
$\chi_2^0 \to \ell_i^\pm\, \ell_j^\mp\, \chi_1^0$.

Our analysis~\cite{Abada:2010kj} 
is focused on how the confrontation of slepton mass splittings
and of low-energy LFV
observables may provide important information about the underlying
mechanism of LFV.

\section{THE SUSY SEESAW}
Our analysis is conducted in the framework of the cMSSM extended by three 
right-handed neutrino superfields, so that the leptonic part of the
superpotential is given by  
$\mathcal{W}^\text{lepton}=\hat N^c\,Y^\nu\,\hat L \, \hat H_2 +
\hat E^c\,Y^l\,\hat L \, \hat H_1 +
\frac{1}{2}\,\hat N^c\,M_N\,\hat N^c$,
where the neutrino Yukawa couplings $Y^{\nu}$ and the
Majorana mass $M_N$ matrix are assumed to be diagonal in flavour space
($Y^l=\operatorname{diag}  (Y^e, Y^\mu,Y^\tau)$, 
$M_N=\operatorname{diag}  (M_{N_i})$ $i=1,2,3$), without loss of generality.
New terms are also added to the soft-SUSY breaking Lagragian, and
universality of the soft-SUSY
breaking parameters is assumed at some high-energy
scale, choosen to be the gauge coupling unification
scale $M_X \sim M_\text{GUT} \sim 10^{16}$ GeV. 
In the so-called seesaw limit, one has the usual seesaw equation for
the light neutrino masses
$m_\nu = - {m_D^\nu}^T M_N^{-1} m_D^\nu $, 
where $m_D^\nu=Y^\nu\,v_2$ and $v_i$ are the vacuum expectation
values of the neutral Higgs scalars, 
$v_{1(2)}= v\,\cos (\sin) \beta$, with $v=174$ GeV.
A convenient means of parametrizing the neutrino Yukawa couplings,
while at the same time allowing to accommodate the experimental data,
is given by the Casas-Ibarra parametrization~\cite{Casas:2001sr},
which reads at the seesaw scale $M_N$
\begin{equation}\label{eq:seesaw:casas}
Y^\nu v_2=m_D^\nu \,=\, i \sqrt{M^\text{diag}_N}\, R \,
\sqrt{m^\text{diag}_\nu}\,  {U^\text{MNS}}^{\dagger}\,.
\end{equation}
In Eq.~(\ref{eq:seesaw:casas})
$R$ is a $3 \times 3$ complex 
orthogonal matrix (parametrized by 3 complex angles $\theta_i$), that
encodes the possible mixings involving the right-handed neutrinos, in
addition to those of the low-energy sector
(i.e. $U^{\text{MNS}}$). We use the standard parametrization of the 
$U^{\text{MNS}}$, 
with the three mixing angles in the intervals favoured by current
best-fit analyses~\cite{GonzalezGarcia:2010er}.

\subsection{Low-energy LFV observables}
In the presence of mixings in the lepton sector,  
$Y^\nu$ is clearly non-diagonal in flavour space, and 
the running from $M_X$ down to the seesaw  scale will induce flavour mixing
in the otherwise (approximately) flavour conserving SUSY breaking
terms. As an example, at low energies the slepton-doublet soft-breaking mass,
$m_{\tilde{L}}^2$ (which at the GUT scale is $m_{\tilde{L}}^2 =
\operatorname{diag}  (m_0^2)$) now reads
\begin{equation}%%\label{eq:slepton:RGE:LLog}
(m_{\tilde{L}}^2)_{ij} \approx \left(
m_0^2 + 0.5 M_{1/2}^2-  m_0^2 |y| ({Y^l})^2_{ii}  \right)
+ (\Delta m_{\tilde{L}}^2)_{ij},\nonumber
\end{equation}
\begin{equation}\label{eq:slepton:RGE:LLog}
(\Delta m_{\tilde{L}}^2)_{ij} \approx
-\frac{1}{8\, \pi^2} (3 m_0^2+ A_0^2) ({Y^{\nu}}^\dagger 
L Y^{\nu})_{ij}, \end{equation}
where $L_{kl}=\log(M_X/M_{N_k}) \delta_{kl}$.
Hence, slepton mass matrices are non-diagonal,
and there is a misalignement between slepton
mass and interaction eigenstates. 
Notice that in the framework of the SUSY seesaw, there is only one
source of (s)lepton flavour violation: the neutrino Yukawa couplings,
$Y^\nu$. 

Flavour violating transitions can occur
in the charged slepton sector, giving rise to low-energy LFV
observables such as $l_i \to l_j \gamma$, $l_i \to 3
l_j$  and  $\mu-e$ conversion in nuclei.

\subsection{LFV at the LHC}
Within the cMSSM, and in the absence of explicit mixing in the slepton 
sector, there
are only two sources of non-universality for the masses of left- and
right-handed sleptons: 
(i) RGE effects proportional to $({Y^l})^2_{ij}$  (see
Eqs.~(\ref{eq:slepton:RGE:LLog})); (ii) $LR$ mixing effects, also 
proportional to the lepton masses ($ m^l_{i}\ \tan \beta $).
Hence, the cMSSM mass differences between the first two families are 
extremely small implying that, to a large extent, the left- and
right-handed selectrons and smuons are nearly degenerate, the mass
splitting typically lying at the per mille level.

In the SUSY seesaw, the radiative corrections introduced by
the neutrino Yukawa couplings
induce both flavour conserving and flavour violating  contributions to
the slepton soft masses: in addition to generating LFV effects, the
new terms proportional to $Y^\nu$ will also break the approximate
universality of the first two generations. An augmented mixing between
$\tilde e $, $\tilde \mu$ and $\tilde \tau$ translates into larger 
mass splittings for the mass eigenstates. In particular, as 
noticed in~\cite{Buras:2009sg}, large mixings involving the third generation
can lead to sizable values of the mass splitting between slepton mass
eigenstates, while avoiding the stringent BR($\mu \to e \gamma$) constraint.
In the latter case (i.e. large $Y^\nu_{32,33}$), the mass
splittings between left-handed sleptons of 
the first two generations 
are given by 
\begin{eqnarray}
&&\frac{\Delta m_{\tilde \ell}}{m_{\tilde \ell}} (\tilde e_{_L},
\tilde\mu_{_L})
=
\frac{|m_{\tilde e_L}-m_{\tilde \mu_L}|}{<m_{\tilde
    \ell_{e,\mu}}>} \ \ \ \ 
\nonumber \\
&&  \approx \frac{1}{2}\,
\frac{\Delta m_{\tilde \ell}}{m_{\tilde \ell}} (\tilde \mu_{_L}, \tilde
\tau_{_2}) \approx \frac{1}{2}\,\left|
\frac{(\Delta m_{\tilde L}^2)_{_{23}}}{(m_{\tilde L}^2)_{_{33}}}
\right|, 
\end{eqnarray}
and can be sizable, well within the expected sensitivity
of the LHC~\cite{Allanach:2000kt}.
Furthermore, notice that in the framework
of the SUSY seesaw, large slepton mass splittings only emerge for
the left-handed sleptons ($\tilde \mu_R$ and $\tilde e_R$ remain
approximately degenerate).

In the cMSSM, the decays of the $\chi_2^0$ into a di-lepton final
state $\chi_2^0 \to \ell_i^\pm \,\ell_i^\mp\, \chi_1^0$ are flavour
conserving, implying that if measurable, the kinematical edges of a
di-lepton mass distribution, $m_{\ell_i \ell_i}$ necessarily lead to
the reconstruction of intermediate sleptons of the same flavour,
$\tilde \ell^i_{L,R}$.

SUSY models violating strict lepton flavour symmetry may leave
distinct imprints on the di-lepton mass distribution, depending on
whether the soft-breaking slepton terms are non-universal (but flavour
conserving) or truly flavour-violating.  In the first case, the most
significant effect will be a visible displacement of the kinematical
edges in each 
of the di-lepton distributions: for instance, the edge corresponding
to $\tilde e_L$ in $m_{ee}$ will not appear at the same values as that
of $\tilde \mu_L$ in $m_{\mu \mu}$, implying that $m_{\tilde e_L}
\neq m_{\tilde \mu_L}$.
The second case will lead to far richer imprints: 
in addition to a relative displacement of the $\tilde \ell^i_X$ in the
corresponding $m_{\ell_i \ell_i}$ distributions, the
most striking effect is the appearance of new edges in a
di-lepton mass distribution: provided there is a large flavour mixing
in the mass eigenstates (and that all the decays are kinematically
viable), one can have 
$\chi_2^0 \to %\left
\{
%\begin{array}{l}
\tilde \ell^i_L\, \ell_i, 
%\vspace*{1mm}\\
\tilde \ell^i_R\,\ell_i ,
%\vspace*{1mm}\\
\tilde \ell^j_X \,\ell_i
%\end{array}
%\right
\}
\to \chi_1^0 \,\ell_i\, \ell_i $
so that in addition to the two $\tilde \ell_{L,R}^i$ edges, 
a new one would appear due to the exchange of $\tilde
\ell_{X}^j$.

Having a unique source of flavour violation implies that the
high-energy LFV observables, 
${\Delta m_{\tilde \ell}}/{m_{\tilde \ell}} (\tilde \ell_i, \tilde 
\ell_j)$, are strongly correlated with the low-energy ones (BRs and
CR). In the absence of a direct means of testing the SUSY seesaw, the
interplay of these sets of observables may allow to either strengthen
the seesaw hypothesis, or even disfavour the seesaw (thus suggesting
additional or even distinct sources of flavour violation).

\section{RESULTS}

\subsection{Slepton mass reconstruction}
The identification of the several high-energy LFV observables at the
LHC implies that several requirements must be met.
First of all, the 
spectrum must be such that the decay chain 
$ \chi_2^0\to \tilde  \ell  \ell \to \chi_1^0 \ell \ell$, with intermediate
real (on-shell) sleptons, is allowed; secondly, the outgoing leptons
should be sufficiently hard: $m_{\chi_2^0}-m_{\tilde  \ell_L, \tilde
  \tau_2} > 10$ GeV; moreover, the $\chi_2^0$ production rates, and
the BR($\chi_2^0\to \chi_1^0 \ell \ell$) must be large enough to
ensure that a significant number of events is likely to be observed at
the LHC.  
 
The above requirements impose strong constraints on the cMSSM
parameters (i.e. on $m_0$, $M_{1/2}$, $A_0$ and $\tan \beta$). 
In~\cite{Abada:2010kj}, the cMSSM parameter space was thouroughly
investigated, leading to the identification of regions where the slepton
masses could be successfully reconstructed. This study allowed to
identify several benchmark points (to which two LHC ones were added),
which were used in the numerical studies.

\begin{table}
{\footnotesize
\begin{tabular}{|c|c|c|c|c|}
\hline
 \ \ \ \ Point \ \ \ \ &  \ \ $m_0$\ \ &
\ \   $M_{1/2}$\ \  &\ \ 
$A_0$ \ \  & \ \  $\tan \beta$\ \ \\
\hline
P1& 110 & 528& 0& 10\\
\hline 
P2& 110& 471& 1000& 10\\
\hline 
P3 & 137& 435 &  -1000 &  10\\
\hline 
P4& 490& 1161& 0& 40\\
\hline 
CMS-HM1& 180&850 &0 &10 \\
\hline 
ATLAS-SU1& 70& 350& 0& 10\\
\hline 
\end{tabular}
\\
}\caption{mSUGRA benchmark points selected for the  LFV
    analysis: $m_0$, $M_{1/2}$ (in GeV) and $A_0$ (in TeV), and $\tan
    \beta$ ($\mu >0$). HM1 and SU1 are
    LHC CMS- and ATLAS-proposed benchmark points~\cite{LHC:points}.
}\label{table:cmssm:di-lepton:values} 
\end{table}

If such events are indeed observable, and successfully reconstructed, 
one expects a 0.1\% precision in the measurement of the kinematical
edges of the di-lepton invariant mass
distributions~\cite{Hinchliffe:1996iu,Bachacou:1999zb}, 
$m_{\ell \ell}=\frac{1}{m_{\tilde{\ell}}} \sqrt{ 
(m^2_{{\chi}^0_2} -m^2_{\tilde{\ell}}) (
  m^2_{\tilde{\ell}}-m^2_{{\chi}^0_1})}$. In turn, this will allow to
infer the slepton mass differences with a precision of $\sim 10^{-4}$
for a $\tilde e - \tilde \mu$ relative mass 
difference~\cite{Allanach:2000kt}; in our analysis we adopted 
a more conservative view, assuming maximal
sensitivities of $\mathcal{O}(0.1\%)$ for $\Delta m_{\tilde
\ell}/m_{\tilde \ell} (\tilde e ,\tilde \mu)$ and $\mathcal{O}(1\%)$
for $\Delta m_{\tilde \ell}/m_{\tilde \ell} (\tilde \mu , \tilde
\tau)$.

\subsection{LFV at low and high energies}
In the framework of the cMSSM (without a seesaw mechanism), the
dilepton invariant mass distribution ($\ell=e,\mu$) for the study
points of Table~\ref{table:cmssm:di-lepton:values} leads to double
triangular distributions (except for point P1), with superimposed edges
corresponding to the exchange of left- and right-handed selectrons and
smuons~\cite{Abada:2010kj}. The numerical scans of the
parameter space further confirm that in the cMSSM both $
\Delta  m_{\tilde{\ell}}/m_{\tilde{\ell}} 
({\tilde e_L},{\tilde \mu_L})$ and 
$\Delta  m_{\tilde{\ell}}/m_{\tilde{\ell}} ({\tilde e_R},{\tilde
  \mu_R})$ lie in the range $10^{-7} - 10^{-3}$. 

Even in the most minimal implementation of the seesaw,
assuming that all flavour mixing in $Y^\nu$ is only stemming from the 
$U^\text{MNS}$ mixing matrix (i.e. taking the conservative limit
$R=1$ in Eq.~\ref{eq:seesaw:casas}), the left-handed slepton mass
splittings are much larger than in the cMSSM, with  values as large as 
$\mathcal{O}(10\%)$. 
Very large splittings are associated with heavy seesaw scales (in
particular, $M_{N_3}$) and/or large, negative values of $A_0$. 
Aside from the perturbativity bounds on $Y^\nu$, the most important
constraints on the seesaw parameters arise from the
non-observation of LFV processes: 
since both flavour violating BRs and slepton mass splittings 
originate from the same unique source ($Y^\nu$), compatibility with
current bounds, in particular on BR($\mu \to e \gamma $) and BR($\tau
\to \mu \gamma$), may preclude sizable values for the slepton mass
splittings\footnote{This is in contrast with other scenarios 
of (effective) flavour
violation in the slepton sector where the
different off-diagonal elements of the slepton mass matrix can be
independently varied~\cite{Buras:2009sg}.}.
This unique synergy is instrumental in devising a strategy to test
the SUSY seesaw via the interplay of high- and low-energy observables.

\begin{figure}[ht!]
\begin{center}
\begin{tabular}{l}
\epsfig{file=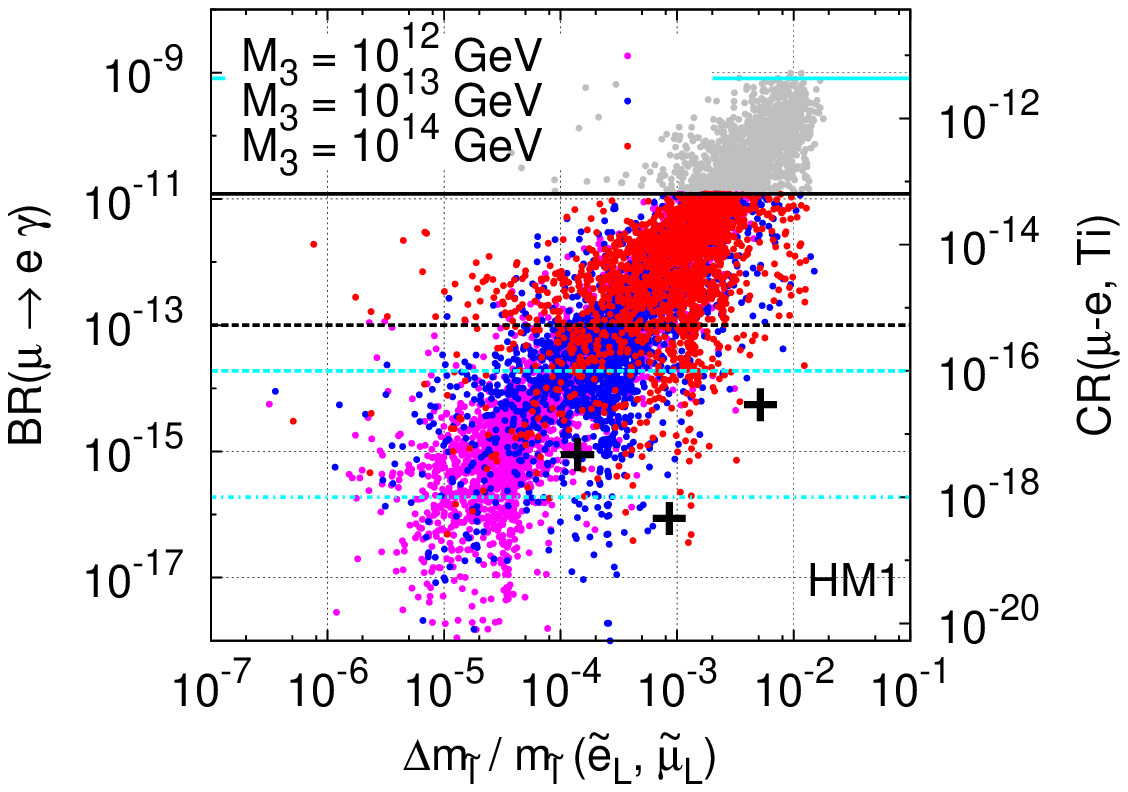, width=70mm}
\\
\epsfig{file=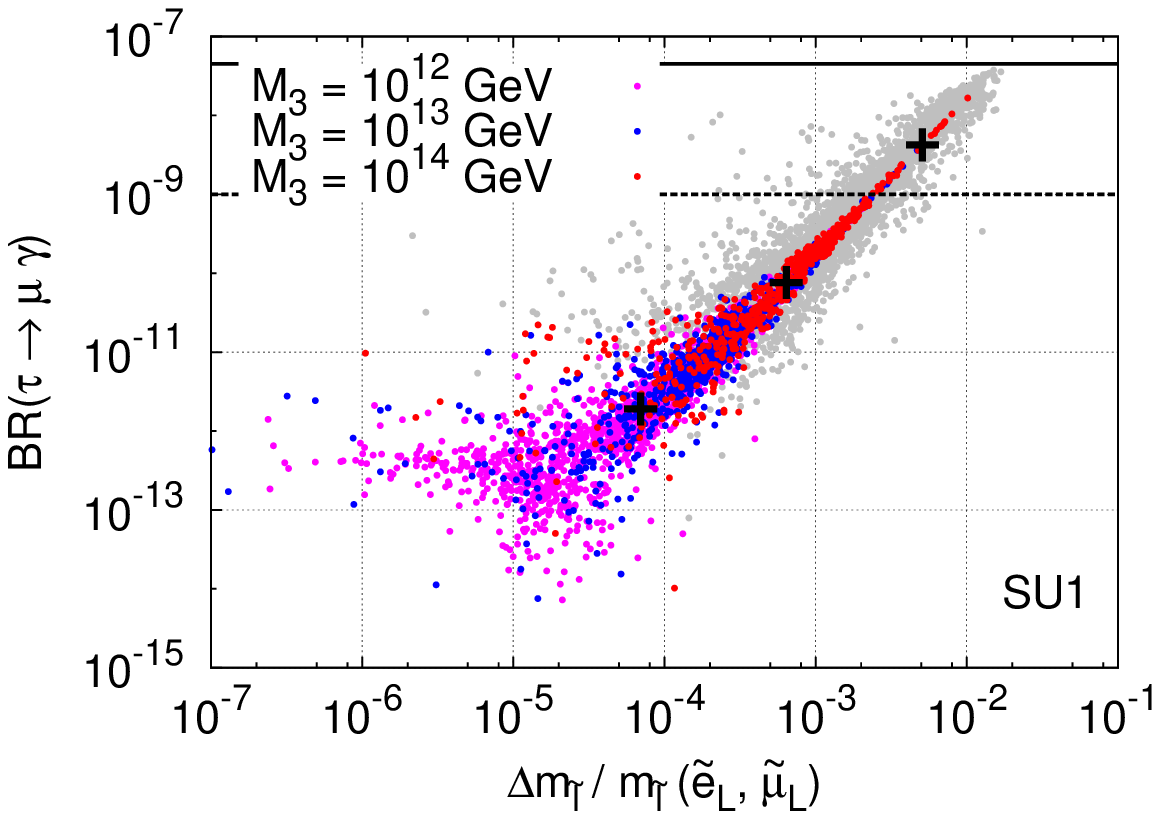, width=65mm}\end{tabular}
\caption{
Upper panel: BR($\mu \to e \gamma$) as a function of the
$\tilde e_L - \tilde \mu_L$ mass difference, for seesaw variations of
point HM1. On the secondary right y-axis, the corresponding
predictions of CR($\mu-e$, Ti). Lower pannel: BR($\tau \to \mu
\gamma$) as a function of the
$\tilde e_L - \tilde \mu_L$ mass difference, for seesaw variations of
point SU1.
Horizontal lines denote the corresponding current bounds/future
sensitivities.  The distinct coloured regions correspond to three
different values of $M_{N_3}=\{ 10^{12},\ 10^{13}, \ 10^{14}\}$ GeV.
We take the Chooz angle to be $\theta_{13}=0.1^\circ$, and 
the remaining parameters were set as $M_{N_1}=10^{10}$ GeV,
$M_{N_2}=10^{11}$ GeV, with the complex $R$
matrix angles being randomly varied as $|\theta_i| \in [0, \pi]$,
and $\arg(\theta_i) \in [-\pi, \pi]$. 
}\label{fig:BRCR.MS:Rcomplex:M3:theta13}
\end{center}
\end{figure}

If the LHC measures a
given mass splitting, predictions can be made regarding the associated
LFV BRs (for an already reconstructed set of mSUGRA
parameters). Comparison with current bounds (or possibly an already
existing BR measurement) may allow to derive some hints on the
underlying source of flavour violation: a measurement of a slepton mass
splitting of a few percent, together with a measurement of a
low-energy observable (in agreement to what
could be expected from the already reconstructed SUSY spectrum) would
constitute two signals of LFV that could be simultaneously explained through
one common origin - a type-I seesaw mechanism. 
On the other hand, two conflicting situations may occur:
(i) a measurement of a mass splitting associated
to LFV decays experimentally excluded; 
(ii) observation of LFV low-energy
signal, and (for an already reconstructed SUSY spectrum) approximate
slepton mass universality.
These scenarios would either suggest that
non-universal slepton masses or low-energy LFV 
would be due to a mechanism other than 
such a simple realisation of a type-I seesaw (barring accidental 
cancellations or different neutrino mass schemes). 
For instance, 
a simple explanation for the first scenario would be that
the mechanism for SUSY breaking is
slightly non-universal (albeit flavour conserving).  

To illustrate this interplay, we conduct a general
scan over the seesaw parameter space. 
In Fig.~\ref{fig:BRCR.MS:Rcomplex:M3:theta13} we display different
low-energy LFV observables as a function of the $\tilde e_L - \tilde
\mu_L$ mass difference, and choose for this overview of the 
SUSY seesaw the LHC points HM1 and SU1.

\begin{figure*}%[ht!]
\begin{center}\vspace*{-8mm}
\begin{tabular}{l}
\epsfig{file=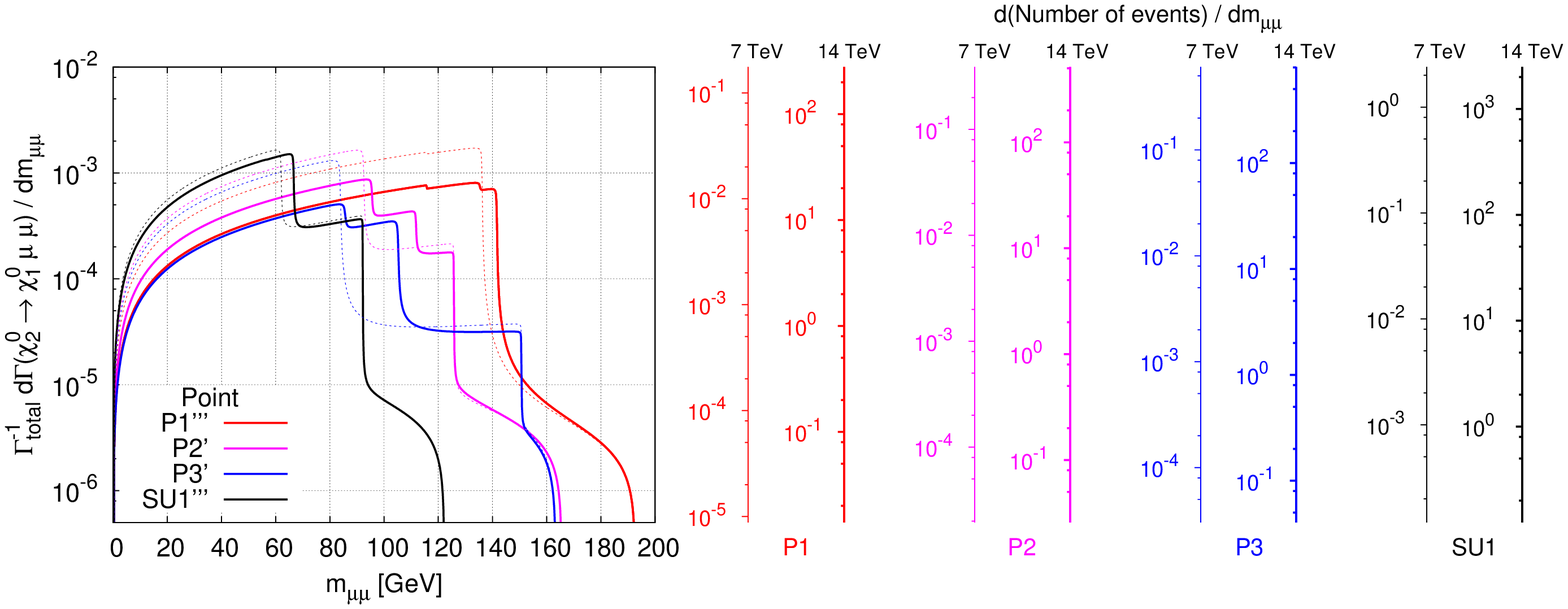, width=100mm}\vspace*{-2mm}
\\
\epsfig{file=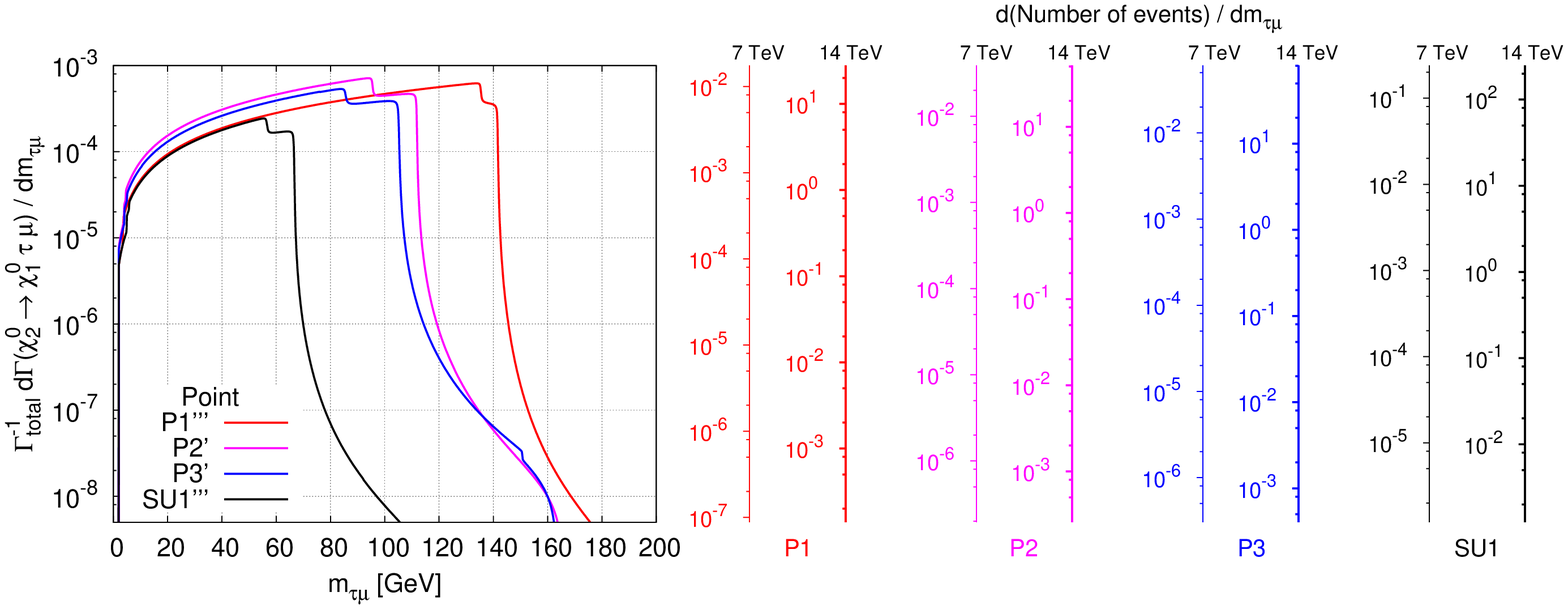, width=100mm}
\vspace*{-8mm}
\end{tabular}
\caption{Upper panel:
BR($\chi_2^0 \to \mu \mu \chi_1^0$) as a function of the
di-muon invariant mass $m_{\mu \mu}$ (in GeV), with dotted lines coresponding 
to the corresponding cMSSM case; lower panel:  
BR($\chi_2^0 \to \mu \tau \chi_1^0$) as a function of the
di-lepton invariant mass $m_{\tau \mu}$ (in GeV). 
We consider different realisations of SUSY seesaw points:
P1$^{\prime \prime \prime}$ (red), 
P2$^{\prime}$ (pink), P3$^{\prime}$ (blue) and 
SU1$^{\prime \prime \prime}$ (black). 
Secondary y-axes denote the corresponding expected number of events for 
$\sqrt s = 7$  TeV and 14 TeV, respectively with 
$\mathcal{L}=1\ \text{fb}^{-1}$ and $\mathcal{L}=100\ \text{fb}^{-1}$.}
\label{fig:edges:seesaw:cMSSM}
\end{center}
\end{figure*}

As can be seen from the upper panel of
Fig.~\ref{fig:BRCR.MS:Rcomplex:M3:theta13}, if a SUSY type-I seesaw is indeed
at work, and $\theta_{13}$ has been constrained to be extremely small,
a measurement of $\Delta m_{\tilde \ell}/m_{\tilde \ell} (\tilde e_L ,
\tilde \mu_L)$ between $0.1\%$ and 1\%, in association with a
reconstructed sparticle spectrum similar to HM1, would be
accompanied (with a significant probability) by the observation of
BR($\mu \to e \gamma$) at MEG~\cite{Ritt:2006cg} 
(and we notice here that, even for very large
values of $M_{N_3}$, the constraints on the parameter space from
BR($\mu \to e \gamma$) would preclude the observation of a $\tau \to \mu
\gamma$ transition for an HM1-like spectrum).  

Although LHC production prospects have to be taken into account, 
when compared to HM1, SU1 offers  a less promising framework  for
the observation of sizable mass splittings at the LHC (unless a
precision of around $10^{-3}$ for $\Delta m_{\tilde \ell}/m_{\tilde
\ell} (\tilde e_L , \tilde \mu_L)$ can indeed be achieved). 
However the most interesting lepton
flavour signature of SU1 is related to its potential to induce
large BR($\tau \to \mu \gamma$), within the future sensitivity of
SuperB~\cite{Bona:2007qt}: a measurement of 
$\Delta m_{\tilde \ell}/m_{\tilde \ell} (\tilde e_L , \tilde \mu_L)$ 
$\sim 0.1\% - 1\%$ at the LHC would imply
BR($\tau \to \mu \gamma$)$\gtrsim 10^{-9}$, and would hint towards a
heavy seesaw scale, $M_{N_3} \gtrsim 10^{13}$ GeV.

\subsection{LFV at the LHC: $\chi_2^0$ decays}
In the upper panel of 
Fig.~\ref{fig:edges:seesaw:cMSSM}, we display the BR($\chi_2^0 \to
\mu \,\mu \,\chi_1^0$) as a function of the di-muon invariant mass $m_{\mu
\mu}$ for different SUSY seesaw points (see~\cite{Abada:2010kj}), 
comparing the distributions with those of the cMSSM.
As is manifest from Fig.~\ref{fig:edges:seesaw:cMSSM}, the impact of
the seesaw at the level of the di-muon mass distributions is quite
spectacular, particularly in the appearance of a third edge in most
of the benchmark points considered. With the exception of the seesaw 
variation of point
P1, all other distributions now exhibit the edge corresponding
to the presence of an intermediate $\tilde \tau_2$, implying that the decay
occurs via $\chi_2^0 \to \tilde \tau_2 \mu \to  \mu\, \mu\, \chi_1^0$. 
For instance, for point P2$^{\prime}$, the BR($\chi_2^0 \to \mu \mu
\chi_1^0$) via intermediate $\tilde \mu_L$, $\tilde \mu_R$ and $\tilde
\tau_2$  are  2.6\%, 1.1\% and 1.6\%, respectively. 
Interestingly, the fact that the SUSY seesaw leads to increased mass
splittings only for the left-handed sleptons
might provide another potential fingerprint for this mechanism of
LFV. Compiling all the data collected throughout our numerical analysis, we
have found that the maximal splitting between right-handed smuons and
selectrons is 
$\left.\frac{\Delta m_{\tilde \ell}}{m_{\tilde \ell}} (\tilde \mu_R, \tilde
e_R)\right|_\text{max}\, \approx \, 0.09 \%$ (while
within the SUSY seesaw $\Delta m_{\tilde \ell}/m_{\tilde \ell} (
\tilde \mu_L, \tilde e_L) $ could easily reach values of a few \%). 
Should the LHC measure mass splittings between right-handed sleptons
of the first two families that are significantly above the 0.1\%
level, this could provide important indication that
another mechanism of FV should be at work: among the many possibilities, 
a likely hypothesis would be the non-universality of the slepton
soft-breaking terms. 

Finally, we display in the lower panel of 
Fig.~\ref{fig:edges:seesaw:cMSSM}
the prospects for direct flavour violation in
$\chi_2^0$ decays: in addition to the possibility of having staus in
the intermediate states, one can also have opposite-sign, different
flavour final state di-leptons.
In particular, one can have $\chi_2^0 \to \mu \tau \chi_1^0$, with a
non-negligible associated branching ratio. For $\sqrt{s} = 14$ TeV and
$\mathcal{L}=100\ \text{fb}^{-1}$, the expected number of events
(without background analysis nor detector simulation) is
$\mathcal{O} (10^3)$.

\section{CONCLUDING REMARKS}
We have discussed that if the seesaw is indeed the source of
both neutrino masses and leptonic mixings, and also accounts for low-energy
LFV observables within future sensitivity reach, interesting slepton
phenomena are expected to be observed at the LHC: in addition to the
mass splittings, the most striking effect will be the possible
appearance of new edges in di-lepton mass distributions.  From the
comparison of the predictions for the two sets of observables (high
and low energy) with the current experimental bounds and future
sensitivities, one can either derive information about the otherwise
unreachable seesaw parameters, or disfavour the type-I SUSY seesaw as
the unique source of LFV. A complete analysis can be found in
Ref.~\cite{Abada:2010kj}.

\end{document}